\documentclass[reprint,amsmath,amssymb,aps]{revtex4-1}
\usepackage{dcolumn}
\usepackage{bm}
\usepackage{graphicx}
\usepackage{amssymb}
\usepackage{undertilde}
\begin{document}
\title{Coherent-state optical qudit cluster state generation and teleportation via homodyne detection}
\author{Jaewan Kim}
 \email{jaewan@kias.re.kr}
\author{Juhui Lee}%
\author{Se-Wan Ji}
\affiliation{
School of Computational Sciences, Korea Institute for Advanced Study, Hoegiro 87, Dongdaemun, Seoul 130-722, Korea}

\author{Hyunchul Nha}
\affiliation{Department of Physics, Texas A$\&$M University at Qatar, Doha, Qatar}

\author{Petr M. Anisimov}
\author{Jonathan P. Dowling}
\affiliation{Hearne Institute for Theoretical Physics and Department of Physics and Astronomy\\Louisiana State University, Baton Rouge, LA 70803, U.S.A.}
\date{\today}
\begin{abstract}
Defining a computational basis of pseudo-number states, we interpret a coherent state of large amplitude, $|\alpha|\gg\frac{d}{2\pi}$, as a qudit --- a $d$-level quantum system --- in a state that is an even superposition of $d$ pseudo-number states. A pair of such coherent-state qudits can be prepared in maximally entangled state by generalized Controlled-$Z$ operation that is based on cross-Kerr nonlinearity, which can be weak for large $d$. Hence, a coherent-state optical qudit cluster state can be prepared by repetitive application of the generalized Controlled-$Z$ operation to a set of coherent states. We thus propose an optical qudit teleportation as a simple demonstration of cluster state quantum computation.
\end{abstract}
\pacs{Valid PACS appear here}
\maketitle
Quantum computation is expected to speed up some computational problems exponentially and some others quadratically compared to the best known digital computation~\cite{NiCh}. Even though many experimental proposals of quantum computers are made, there seem to be many obstacles such as decoherence, scalability, inaccurate operation, and so on~\cite{HuDo}. There are two approaches to the quantum computing --- one that {\em molds} quantum state while the other {\em sculptures} it. Molding of quantum states lies in the heart of original schemes for quantum computers that are based on quantum circuits. In these schemes, one prepares an initial quantum state made of many qubits and applies quantum operations on it, which are followed by a measurement that leads to the result.

Raussendorf and Briegel proposed a special quantum entanglement called a cluster state~\cite{RaBr} and went on proposing cluster state quantum computation with Browne~\cite{RaBrBr}: you prepare a cluster state, a giant maximally entangled state of many qubits, and just measure each qubit away feedforwardly which means measurements are done based on previous measurement results--- effectively sculpturing the state. To make a quantum cluster state, prepare qubits of even superposition of computational basis kets $\left| 0 \right\rangle$ and $\left| 1 \right\rangle$  at each lattice points and apply $CZ_{\rm ct}$ (c is an index for the control qubit and t is for the target qubit) operations on all neighboring qubits in the lattice.

Even though the number of required qubits is polynomially larger than quantum circuit model, cluster state quantum computation is simpler since only single qubit measurements are needed once a cluster state is prepared.

Based on Knill, Laflamme, and Milburn's all-optical quantum computing~\cite{KnLaMi} and Raussendorf, Browne, and Briegel's cluster state quantum computing~\cite{RaBrBr}, Nielsen and Dawson proposed optical cluster state quantum computing~\cite{Ni, NiDa}. One important demerit of the proposal might be the probabilistic nature of linear optical gating. Nonlinear optics can be used to generate quantum optical entanglements~\cite{SaMi} and generation of optical qubit cluster states are proposed~\cite{Pa, Wa, NgJK}. These schemes, however, need impractically large nonlinearities. Instead of qubits with two basis states, quantum computation using cluster states of  $d$-state quantum systems or qudits has been proposed with the possibility of realization with high-dimensional “Ising” model~\cite{Zh}.

Here we propose a simple deterministic optical scheme to generate a cluster state of qudits. First we notice that the infinite Taylor series of an exponential function can be decomposed into $d$ infinite partial sums each of which asymptotically approaches $e^x/d$ for any finite integer $d$ as can be seen in the following	
\begin{equation}
e^x  = \sum\limits_{k = 0}^{d - 1} {f_k (x)\,{\text{  with  }}f_k (x) = \sum\limits_{m = 0}^\infty  {\frac{{x^{k + md} }}
{{(k + md)!}}} } 
\end{equation}
where
\begin{equation}
\mathop {\lim }\limits_{x \to \infty } \frac{{f_k \left( x \right)}}{{e^x }} = \frac{1}{d}{\;\;\;\rm{ for }\;\;\;}k = 0,\,...\,,d - 1.
\end{equation}

In a similar manner a coherent state $\left| \alpha  \right\rangle$ can be interpreted as a qudit that is evenly superposed in a computational basis with $d$ basis ket vectors when $ \left| \alpha  \right|  \gg \frac{d}{{2\pi }}$ and this condition is assumed throughout this paper.

\begin{equation}
\left| \alpha  \right\rangle  = e^{ - \frac{{\left| \alpha  \right|^2 }}
{2}} \sum\limits_{n = 0}^\infty  {\frac{{\alpha ^n }}
{{\sqrt {n!} }}} \left| n \right\rangle \,\, = \,\frac{1}
{{\sqrt d }}\sum\limits_{k = 0}^{d - 1} {\left|\utilde{k} \right\rangle } 
\end{equation}
with orthonormalized computational basis kets
\begin{equation}
\left| \utilde{k} \right\rangle  = \sqrt d \: e^{ - \frac{{\left| \alpha  \right|^2 }}
{2}} \sum\limits_{m = 0}^\infty  \! {\frac{{\alpha ^{k + md} }\left| {k + md} \right\rangle }
{{\sqrt {(k + md)!} }}} {\;\rm{ for }\;}k = 0,\,...\,,d - 1
\end{equation}
that we call pseudo-number states since each ket is made of photon number states with definite modulo-$d$ number of photons.

By applying a generalized Hadamard transformation $\hat H$ 
\begin{equation}
\hat{H} = \frac{1}{\sqrt{d}} \sum_{k=0}^{d-1}\sum_{l=0}^{d-1}\omega^{kl}\left|\utilde{k}\right\rangle
\left\langle\utilde{l}\right| 
\end{equation}
on computational basis ket $\left| \utilde{k} \right\rangle 
$'s, we can get conjugated basis kets

\begin{equation}
\left| {\widetilde l} \right\rangle   = \hat{H} \left| \utilde{l} \right\rangle = \frac{1}
{{\sqrt d }}\sum\limits_{k = 0}^{d - 1} {\omega ^{lk} } \left| \utilde{k} \right\rangle {\;\;\;\rm{ for }\;\;\;}l = 0,\,...\,,d - 1.
\end{equation}
with $\omega  = e^{\frac{{2\pi i}}{d}} $.

These conjugated basis kets are nothing but the coherent states
\[
\left| {\widetilde l} \right\rangle =\left| {\omega ^l \alpha } \right\rangle 
\]
that we call pseudo-phase states since each ket of this basis is a coherent state centered at a definite optical phase.

A generalized $\hat{Z}$ operator for qudits is defined as
\[
\hat{Z} = \sum\limits_{k = 0}^{d - 1} {\omega ^k } \left| \utilde{k} \right\rangle \left\langle \utilde{k} \right|
=\omega^{\hat{n}}
\]
with a photon number operator $\hat{n}$ and a generalized Controlled-$Z$ operator, $\hat{Z}_{\rm ct}$, is defined as 
\[
\hat{Z}_{{\text{ct}}}  = \sum\limits_{l = 0}^{d - 1} {\left| \utilde{l} \right\rangle _{{\text{c  c}}} \left\langle \utilde{l} \right|}  \otimes \hat{Z}_{\text{t}} ^l
=\omega^{\hat n_{\text c} \hat{n}_{\text t}}
\]
with c and t for control and target qudits respectively.

A generalized $\hat{Z}$ operator can be easily implemented by a phase shifter $e^{\frac{{2\pi i}}{d}\hat n} $ with photon number operator $n$, and a generalized Controlled-$Z$ operator, $\hat{Z}_{\rm ct}$, can be realized by cross-Kerr medium. If the cross-Kerr interaction with Hamiltonian $ H =  - \hbar \chi \hat n_1 \hat n_2 $ is applied to two-coherent-state input $ \left| \alpha  \right\rangle _1 \left| \alpha  \right\rangle _2 $
 for time  $t = \frac{{2\pi }}{{d\chi }}$, we can get
\begin{equation}
\begin{array}{ll}
  & e^{\frac{{2\pi i}}
{d}\hat n_1 \hat n_2 } \left| \alpha  \right\rangle _1 \left| \alpha  \right\rangle _2  \\
=& \hat{Z}_{12} \left( {\frac{1}
{{\sqrt d }}\sum\limits_{k = 0}^{d - 1} {\left| \utilde{k} \right\rangle } } \right)_1 \left( {\frac{1}
{{\sqrt d }}\sum\limits_{l = 0}^{d - 1} {\left| \utilde{l} \right\rangle } } \right)_2  \hfill \\
=& \frac{1}{d}\sum\limits_{k = 0}^{d - 1} {\sum\limits_{l = 0}^{d - 1} {\omega ^{kl} } \left| \utilde{k} \right\rangle _1 \left| \utilde{l} \right\rangle _2 }  \hfill \\
=&\frac{1} {{\sqrt d }}\sum\limits_{k = 0}^{d - 1} {\left| \utilde{k} \right\rangle _1 \left| {\widetilde k} \right\rangle _2 } \,\,{\text{or }}\frac{1}
{{\sqrt d }}\sum\limits_{k = 0}^{d - 1} {\left| {\widetilde k} \right\rangle _1 \left| \utilde{k} \right\rangle _2 }  \hfill \\ 
\end{array} 
\end{equation}
which is a maximally entangled state of two qudits, that is, we can generate a maximal entanglement of pseudo-phase and pseudo-number states by simply applying cross-Kerr interation on two coherent beams. The larger $d$, the easier the implementation of $\hat{Z}_{\rm ct}$ of qudits is since it can be achieved with smaller $\chi t = \frac{{2\pi }}{d}$. If we apply $\hat{Z}_{\rm ct}$  to all neighboring coherent states as illustrated in Fig.1, we can get a cluster state of qudits 
\[
\prod\limits_{\left\langle {{\text{p,q}}} \right\rangle } \omega^{\hat n_{\text{p}} \hat n_{\text{q}} } \prod\limits_{{\text{r}} \in {\text{lattice}}} {\left| \alpha  \right\rangle _{\text{r}} }  
\]
where $\left\langle {{\text{p,q}}} \right\rangle $ represents neighbors in the lattice. Since all the Controlled-$Z$'s are commuting with each other, the order of the operations is not important. 

It used to be believed that two-qubit operations are the most difficult part and single qubit operations are relatively easier in quantum information processing. Now contrary to this conventional wisdom of qubit processing, Conrolled-$Z$ of two qudits and preparation of cluster states of optical qudits gets easier as the dimension $d$ gets larger. 
A generalized $\hat{X}$ operator can be defined as
\[\hat{X} = \sum\limits_{k = 0}^{d - 1} {\left| \utilde{k-1} \right\rangle } \left\langle \utilde{k} \right|\,\,\,{\text{with }}\,\left| \utilde{-1} \right\rangle  = \left| \utilde{d-1} \right\rangle , \]
is similar to Pegg-Barnett phase operator~\cite{PeBa} and could be called pseudo-phase operator. In pseudo-phase basis it can be written
\[
\hat{X}=\sum_{l=0}^{d-1} \omega^l \left|\widetilde{l}\right\rangle \left\langle\widetilde{l}\right|
\]
and $\hat Z$ can be written
\[
\hat Z= \sum_{k=0}^{d-1} \left|\widetilde{k+1}\right\rangle \left\langle\widetilde{k}\right|
\,{\text{with }}\,\left| \widetilde{d} \right\rangle  = \left| \widetilde{0} \right\rangle\;
\]
and can be called a pseudo-number operator~\cite{PeBa}.
The two operators are related to each other through generalized Hadamard operation and the followings can be readily shown.
\[\hat H\hat Z \hat H^\dagger = \hat X ,\]
\[\hat H^\dagger \hat Z \hat H= \hat X^{-1} , \]
and \[\hat H \hat H \equiv R , \]
where $\hat R$ operation reverses the order of the computational basis with 0 to 0, 1 to $d-1$, 2 to $d-2$, and so on.

Now as a simple demonstration of cluster state quantum computation of optical qudits, we propose a qudit teleportation via homodyne detection. Let us first consider a one-step teleportation as in Fig.2. If qudit state $|\phi\rangle_1 = \sum_{l=0}^{d-1}a_l\left|\utilde l\right\rangle_1$ is entangled with a coherent state $|\alpha\rangle_2$ by $\hat Z_{12}$ and the first qudit is measured in pseudo-phase basis into $\left|\widetilde k\right\rangle_1$, then the second qudit becomes $\hat H \hat Z^{-k} |\phi\rangle$ as can be seen in the following. 
\begin{equation}
\begin{array}{rl}
\hat Z_{12}|\phi\rangle_1 |\alpha\rangle_2 = &  \sum_{l,m} a_l \omega^{\hat n_1 \hat n_2}\left|\utilde l\right\rangle_1 \frac{\left|\utilde m\right\rangle_2}{\sqrt d} \\
= & \sum_l a_l \left|\utilde l\right\rangle_1 \left|\widetilde l\right\rangle_2 \\
\xrightarrow{\text{measured into } \left|\widetilde k\right\rangle_1} & \sum_l a_l \omega^{-lk}\left|\widetilde l \right\rangle_2 \\
= & \hat H \hat Z^{-k} |\phi\rangle_2 .
\end{array}
\end{equation}

A qudit teleportation is a repetition of one-step teleportation. 
Alice has a qudit state $\left| \phi  \right\rangle _1  = \sum\limits_{l = 0}^{d - 1} {a_l \left| \utilde{l} \right\rangle _1 }$, and Bob has $\left| \alpha  \right\rangle _2 \left| \alpha  \right\rangle _3$. Bob applies $Z_{23}$ to prepare a maximally entangled qudits and sends the second qudit to Alice. Now Alice applies $Z_{12}$ on qudits 1 and 2 and measures the first and the second qudits in conjugated basis (coherent states) and gets  
$\left| {\widetilde k} \right\rangle _1   \left| {\widetilde s} \right\rangle _2 $ and informs Bob of the values $k$ and $s$.
\begin{equation}
\begin{array}{rl}
\hat Z_{12} \hat Z_{23} |\phi\rangle_1 |\alpha\rangle_2 |\alpha\rangle_3 = & \sum_{l, m, n} a_l \left|\utilde l\right\rangle_1 \omega^{lm} \frac{\left|\utilde m\right\rangle_2}{\sqrt d} \omega^{mn} \frac{\left|\utilde n\right\rangle_3}{\sqrt d}\\
\xrightarrow{\text{measured into }  \left|\widetilde k\right\rangle_1  \left|\widetilde s\right\rangle_2} & \hat H \hat Z^{-s} \hat H \hat Z^{-k} |\phi\rangle_3\\
= & \hat X^{-s} \hat R \hat Z^{-k} |\phi \rangle_3 .
\end{array}
\end{equation}
Even though Bob can recover $\left| \phi  \right\rangle $ by applying $\hat X^s$, $\hat R$ and $\hat Z^{k}$ operations in order as in Fig.3, just knowing Alice's measurement results $k$ and $s$ might be enough to complete the teleportation without actually applying the operations. 
 
Alice's projective measurement of qudits in pseudo-phase basis, which is the essential part of the above qudit teleportation, can be done by a double-arm homodyne detection. The qudit whose pseudo-phase is to be measured is split by a 50/50 beamspliter and quadrature $X_1$ is measured in one arm and $X_2$ in the other by controling local oscillators for each arm. $X_1$ and $X_2$ will fix the pseudo-phase of the measured qudit as in Fig.4.
If we entangle an optical qudit with a coherent state by Controlled-$Z$, we can measure the qudit in pseudo-number basis by measuring the entangled coherent state in pseudo-phase state as in Fig.5.

Even though new proposals of giant Kerr effects have been made, the present limit of cross-Kerr nonlinearity $\chi t$ is the order of 10$^{-4}$, the dimesion $d$ of qudit is the order of 10$^5$, which means the average photon number $|\alpha|^2$ of coherent-state optical qudit should be the order of 10$^{10}$. Much stronger Kerr nonlinearity of ion strings~\cite{Ro} might be exploited for qudit cluster state quantum computation.

In summary, we have proposed a simple way of generating a cluster state of optical qudits from coherent states. This cluster state could provide a platform for practical large scale quantum computation. As a simple demonstration of qudit cluster state quantum computation, a qudit teleportation scheme is proposed. 

We acknowledge useful discussions with Professors Yoonho Kim, Kisik Kim, Nguyen Ba An, Jungsang Kim, Tae-Gon Noh and Sangkyung Choi. This work was partially supported by the IT R\&D program of MKE/KEIT (KI001789). Petr M. Anisimov and Jonathan P. Dowling would like to acknowledge support from the Foundational Questions Institute, the National Science Foundation, and the Northrop-Grumman Corporation.

\begin{figure}[ch]
\includegraphics[angle=0,scale=1.0,width=\textwidth]{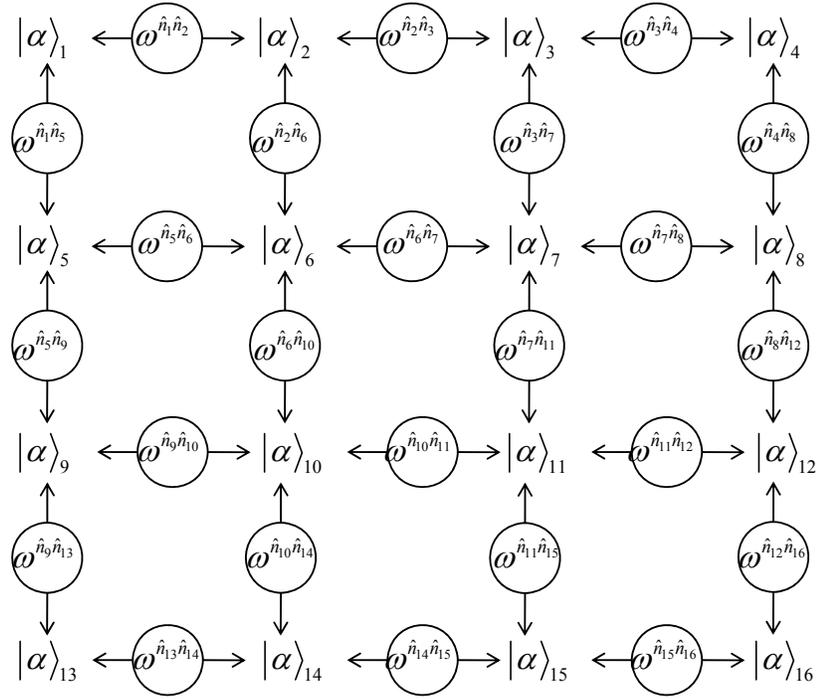}
\caption{Generating a cluster state of coherent-state optical qudits. $\omega^{\hat n_1 \hat n_2}=e^{\frac{2\pi i}{d}{\hat n_1 \hat n_2}}=\hat Z_{12}$ and so on.}
\end{figure}

\begin{figure}[ch]
\includegraphics[angle=0,scale=1.0,width=\textwidth]{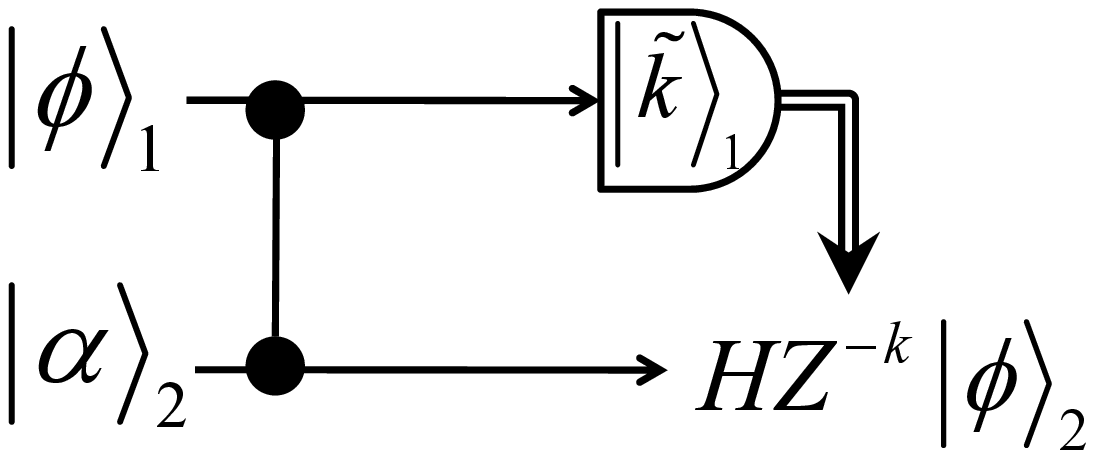}
\caption{One-step teleportation of an optical qudit.}
\end{figure}

\begin{figure}[ch]
\includegraphics[angle=0,scale=1.0,width=\textwidth]{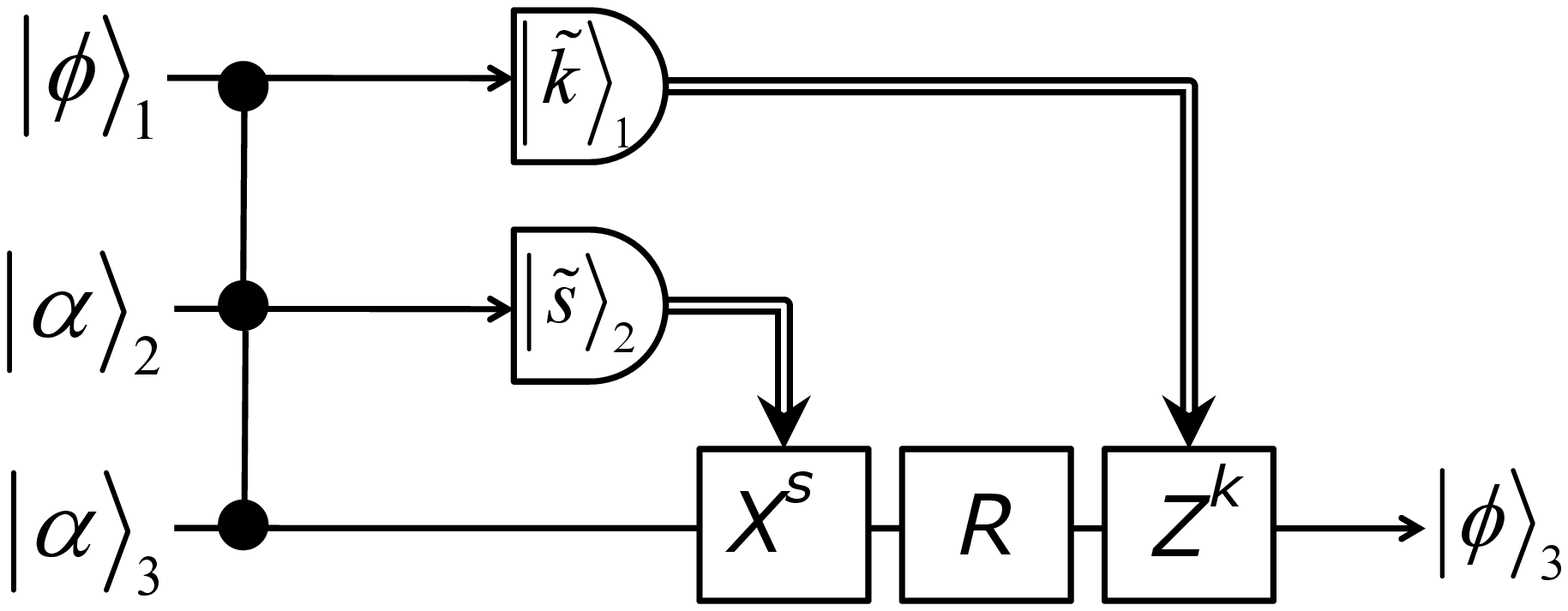}
\caption{Teleportation of an optical qudit.}
\end{figure}

\begin{figure}[ch]
\includegraphics[angle=0,scale=1.0,width=\textwidth]{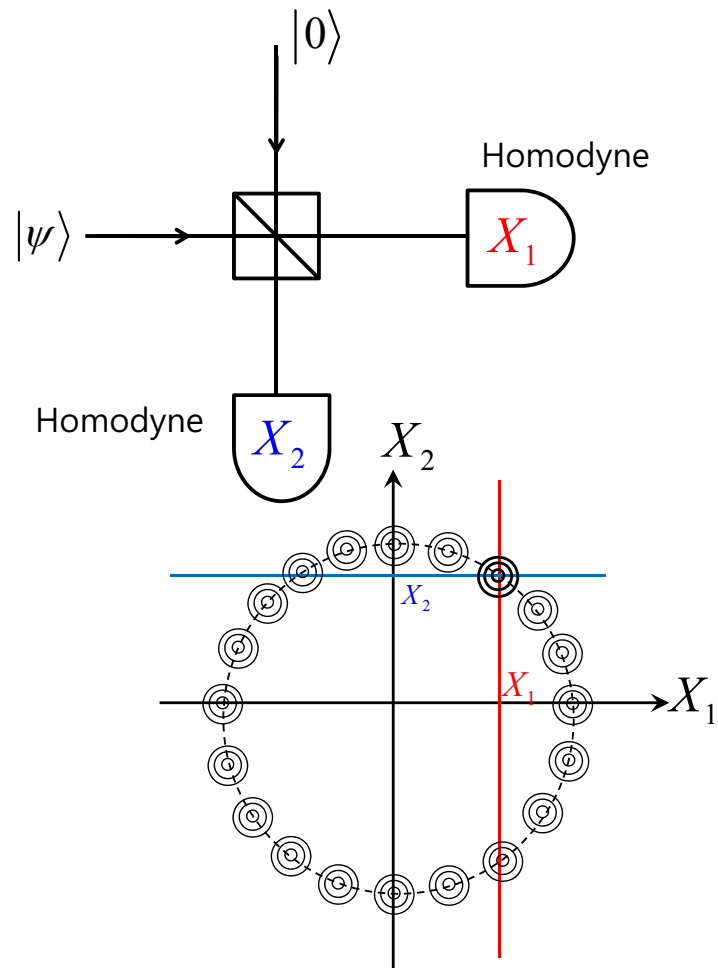}
\caption{Pseudo-phase measurement of an optical qudit.}
\end{figure}

\begin{figure}[ch]
\includegraphics[angle=0,scale=1.0,width=\textwidth]{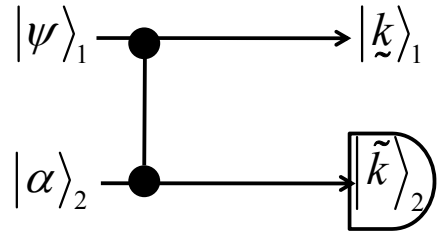}
\caption{Pseudo-number measurement of an optical qudit.}
\end{figure}

\end{document}